\documentclass[preprint,12pt,3p]{elsarticle}

\usepackage{amsmath,amssymb,amsfonts}
\usepackage{mathrsfs}
\usepackage{chemformula}
\usepackage{color}
\usepackage[normalem]{ulem}
\usepackage{stmaryrd}
\usepackage{diagbox}
\usepackage[parfill]{parskip}

 \biboptions{sort&compress}

\newcommand{\blue}[1]{\textcolor{black}{#1}}

\newcommand{\beq}{\begin{equation}}
\newcommand{\eeq}{\end{equation}}
\newcommand{\pfr}[2]{\ensuremath{\frac{\partial #1}{\partial #2}}}
\newcommand{\pfi}[2]{\ensuremath{{\partial #1}/{\partial #2}}}

\newcommand\Pec{\mbox{\textit{Pe}}}

\newcommand\Lew{\mbox{\textit{Le}}}
\newcommand\Dam{\mbox{\textit{Da}}}

\journal{Proceedings of the Combustion Institute}

\begin{document}

\begin{frontmatter}

\title{Three-dimensional diffusive-thermal instability of flames propagating in a plane Poiseuille flow}

\author{Aiden Kelly$^a$, Prabakaran Rajamanickam$^a$, Joel Daou$^a$, Julien R. Landel$^{a,b}$}
\address{$^a$Department of Mathematics, University of Manchester, Manchester M13 9PL, UK \\
$^b$Univ Lyon, Univ Claude Bernard Lyon 1, CNRS, Ecole Centrale de Lyon,\\
INSA Lyon, LMFA, UMR5509, 69622 Villeurbanne, France}

\begin{abstract}
 The three-dimensional diffusive-thermal stability of a two-dimensional flame propagating in a Poiseuille flow is examined. The study explores the effect of three non-dimensional parameters, namely the Lewis number $\Lew$, the Damk{\"o}hler number $\Dam$, and the flow Peclet number $\Pec$.  Wide ranges of   the Lewis number and the flow amplitude are covered, as well as  conditions corresponding to small-scale narrow ($\Dam \ll 1$)  to large-scale wide ($\Dam \gg 1$) channels. The instability experienced by the flame appears as a combination of the traditional diffusive-thermal instability of planar flames and the recently identified instability corresponding to a  transition from symmetric to asymmetric flame. The instability regions are identified in the $\Lew$-$\Pec$ plane for selected values of $\Dam$ by computing the eigenvalues of a linear stability problem. These are complemented by two- and three-dimensional time-dependent simulations describing the full evolution of unstable  flames  into the non-linear regime. In narrow channels, flames are found to be always symmetric about the mid-plane of the channel. Additionally, in these situations, shear flow-induced Taylor dispersion enhances the cellular instability in $\Lew<1$ mixtures and suppresses the oscillatory instability in $\Lew>1$ mixtures. In large-scale channels, however, both the cellular and the oscillatory instabilities are expected to persist.  Here, the flame has a stronger propensity to become asymmetric when the mean flow opposes its propagation and when $\Lew<1$; if the mean flow facilitates the flame propagation, then the flame is likely to remain symmetric about the channel  mid-plane. For $\Lew>1$, both symmetric and asymmetric flames are encountered and are accompanied by temporal oscillations.
\end{abstract}


\end{frontmatter}

\section{Introduction\label{sec:introduction}} 

Flame propagation in a  Poiseuille flow is a classical problem relevant to many practical combustion devices with large and small scales. It provides a suitable framework to investigate fundamental flame-flow interaction phenomena such as flame blow-offs, flashbacks~\cite{lewis2012combustion} and Taylor dispersion effects on flames. The two-dimensional channel configuration offers an attractive platform to investigate such flame phenomena both from the theoretical and experimental viewpoints, see e.g.~\cite{daou2001flame,fernandez2018analysis,pearce2014taylor}.  This platform is also ideal to investigate flame instabilities, including the Darrieus--Landau and diffusive-thermal instabilities as done in~\cite{sarraf2018quantitative,veiga2019experimental,al2019darrieus}. This study will complement the findings of such investigations, by focusing specifically on the effect of the flow scale (channel width) and amplitude on the diffusive-thermal instability.  

\begin{figure}
\centering
\includegraphics[width=0.47\textwidth]{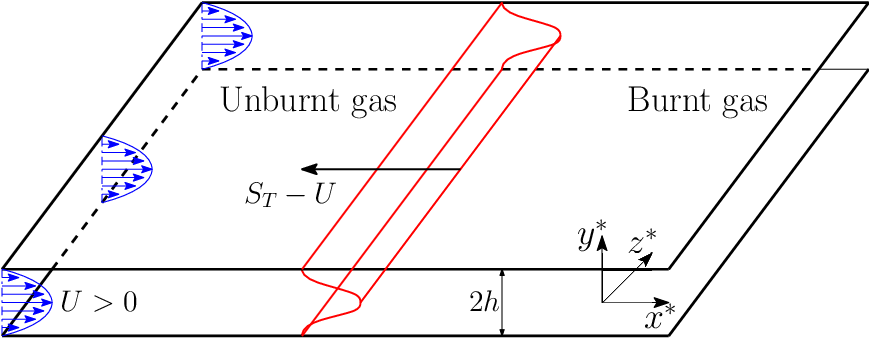}\par \vspace{0.3cm}
\includegraphics[width=0.47\textwidth]{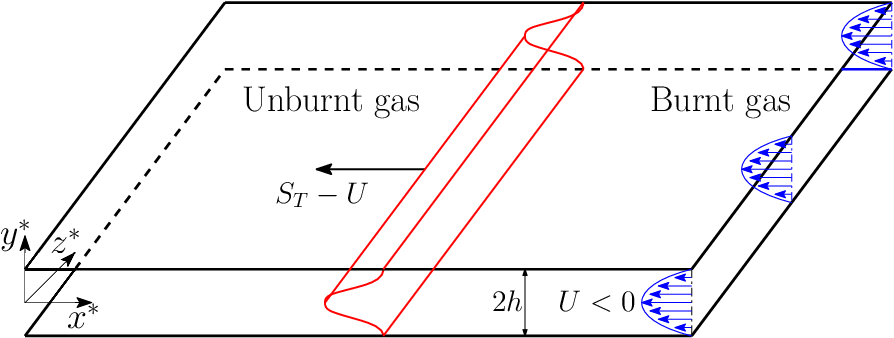}
\caption{Schematic illustration of a premixed flames whose propagation is opposed by a two-dimensional Poiseuille flow  (top) or  aided  by the flow (bottom). The flame propagation speed is $S_T-U$ with respect to the walls, where $S_T$ is the effective burning speed, i.e. the flame speed with respect to the mean flow.} 
\label{fig:setup}
\end{figure}

Figure~\ref{fig:setup} shows a schematic representation of the configuration adopted in our study. Shown are symmetric flames whose propagation is either opposed or aided by the Poiseuille flow.  The flow field is prescribed and its components are given by
\begin{equation}
     v_x^*=\frac{3U}{2} \left(1-\frac{y^{*2}}{h^2}\right), \quad v_y^*=v_z^*=0. \label{Poi}
\end{equation}
Here $x^*$ is the streamwise coordinate, $y^*$   the wall-normal coordinate, $z^*$   the spanwise coordinate, $U$  the mean flow speed, and $h$  the channel half-width. Denoting by $D_T$  the thermal diffusivity and  by $D_F$   the fuel diffusion coefficient,  we can define three relevant non-dimensional parameters, namely the  Lewis number $\Lew$, the flow Peclet number $\Pec$ and the Damk\"{o}hler number $\Dam$ by
\begin{equation} \label{NonDimPar}
   \Lew=\frac{D_T}{D_F},\quad  \Pec= \frac{Uh}{D_T}, \quad \Dam = \frac{h^2/D_T}{\delta_L^2/D_T} ,
\end{equation}
where $\delta_L$ is the laminar flame thickness.  
Large and small values of $\Pec$ here indicate strong and weak  flow intensities,  respectively. Also,  large values of $\Dam$ describe wide channels (or thin  flames) while small  values pertain to narrow  channels (or thick flames).

An investigation of  the diffusive-thermal instability of flames such as the ones in Fig.~\ref{fig:setup} was undertaken by Kurdyumov~\cite{kurdyumov2011lewis} for various values of $\Dam$, $\Lew$ and $\Pec$. The investigation revealed interesting results,
in particular,  it emphasised  the propensity of the flame to become asymmetric (with respect to  $y^*=0$) following  the loss of its initially symmetric  shape for certain values of the parameters. 
It can be noted, however, that the stability analysis in~\cite{kurdyumov2011lewis} is two-dimensional as it considers perturbations independent of the $z^*$-coordinate. In this paper, we shall discard this restriction by considering a three-dimensional analysis. This is important because, as we shall confirm,  the dominant unstable mode turns  out in most cases to correspond  to perturbations depending on the spanwise coordinate $z^*$.  

A stability analysis considering perturbations depending on the spanwise coordinate $z^*$  has been carried out in~\cite{daou2021effect}, but it was limited to thick flames in narrow channels, that is to cases with $\Dam \ll 1$. The main objective of the current paper is to remove the restrictions associated with the perturbations being $z^*$-independent or  $\Dam \ll 1$. This will allow us to provide a more complete characterisation of the flame diffusive-thermal instabilities,
applicable for three-dimensional flames in narrow and wide channels. To this end, we shall adopt a simple constant-density adiabatic model as done in~\cite{kurdyumov2011lewis,daou2021effect}
in order to focus on the diffusive-thermal instability. 
The reader may refer to the literature for information about the  influence of other factors pertinent to  flame propagation in a Poiseuille flow which are not taken into account herein. Such factors include heat loss~\cite{kurdyumov2014propagation,daou2023flame}, equivalence ratio~\cite{sanchez2014effect,fernandez2017effects}, complex chemistry~\cite{fernandez2014differential} and an axisymmetric geometry~\cite{kurdyumov2016structure}.

\section{Governing equations}
\label{sec:basic}
 
 We shall ignore the effects of thermal expansion and heat loss in order to focus on the diffusive-thermal instability, as mentioned above and as done in~\cite{kurdyumov2011lewis,daou2021effect}. Under these conditions, the velocity field may be prescribed and is given by~\eqref{Poi}. As shown in Fig.~\ref{fig:setup}, the flow direction is  from left to right when $U>0$ and from right to left when $U<0$. A two-dimensional symmetric premixed flame, independent of $z^*$, is depicted which  propagates in the negative $x^*$-direction  with constant speed $S_T-U$ with respect to the walls. Here,  $S_T$ is the flame  effective burning speed, that is the flame speed with respect to   the mean flow. To render the flame steady, a reference frame moving with the flame is used  by introducing the coordinate transformation $(x^*,y^*,z^*)\mapsto (x^*+ \blue{S_Tt^*-Ut^*},y^*,z^*)$. In this frame, the velocity field~\eqref{Poi} becomes
\begin{equation}
    v_x^* = \frac{U}{2}\left(1-\frac{3y^{*2}}{h^2}\right) + S_T, \quad v_y^*=v_z^*=0. \nonumber
\end{equation}

\begin{figure*}[h!]
\centering
\includegraphics[width=1\textwidth]{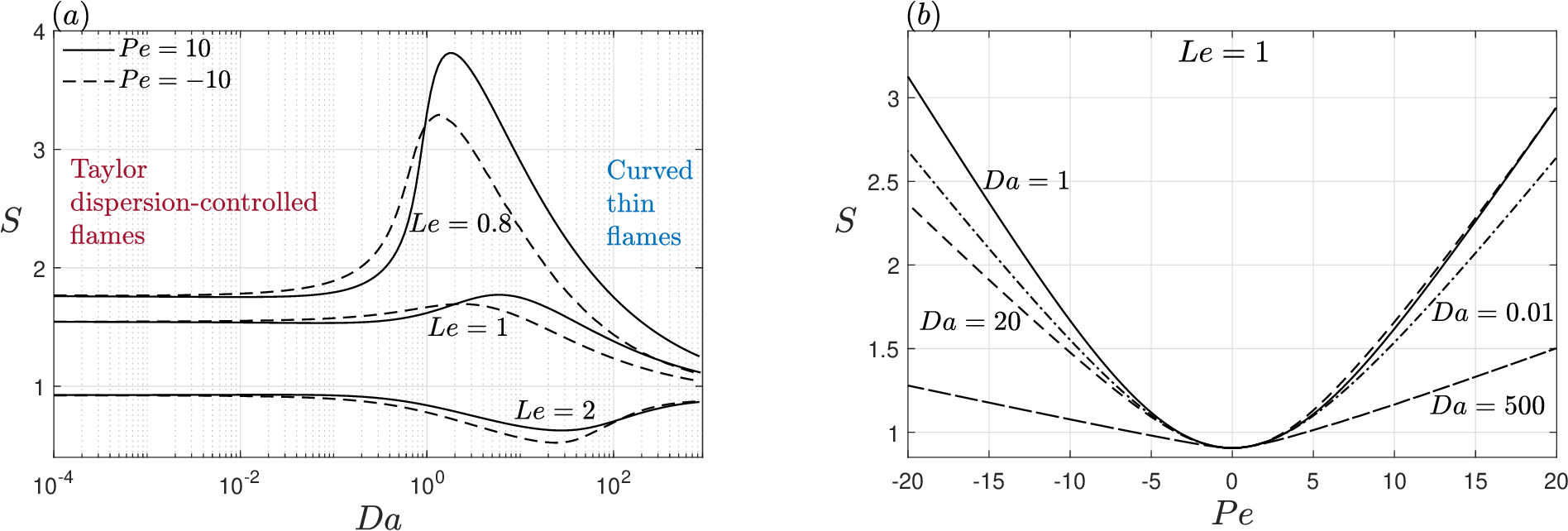}
\vspace{0.05cm}
\caption{The burning speed $S$ computed for selected values of $\Dam$, $\Lew$ and $\Pec$. All computations are performed with $\beta=10$ and $\alpha=0.85$.} \label{fig:S}
\end{figure*}
The analysis assumes fuel-lean conditions for the reactive mixture and adopts a one-step reaction model. The mass of fuel burnt per unit volume and unit time can then be written as $B\rho Y_F e^{-E/RT}$, where $B$ is the pre-exponential factor, $\rho$ is the constant density, $Y_F$ is the fuel mass fraction, $E/R$ is the activation temperature and $T$ is the temperature. We also define the heat-release parameter $\alpha$, the Zeldovich number $\beta$ and the planar laminar burning speed $S_L$ (for $\beta\gg 1$) by the usual  expressions 
\begin{align}
    \alpha = \frac{T_{ad}-T_u}{T_u}, \qquad \beta = \frac{E (T_{ad}-T_u)}{RT_{ad}^2}, \qquad
    S_L = (2\Lew\beta^{-2} B D_T e^{-E/RT_{ad}})^{1/2}.\nonumber
\end{align}
Here, $T_{ad}=T_u+qY_{F,u}/c_p$ is the adiabatic flame temperature, $q$  the amount of heat released per unit mass of fuel burnt, $Y_{F,u}$   the fuel mass fraction in the unburnt mixture, and $c_p$  the constant-pressure specific heat. The laminar flame thickness $\delta_L$  is defined by $\delta_L=D_T/S_L$ and the non-dimensional effective burning speed $S$ by $S=S_T/S_L$.

We adopt the non-dimensionalized variables
\begin{align}
    t = \frac{D_Tt^*}{\delta_L^2}, \quad (x,z)= \frac{1}{\delta_L}(x^*,z^*), \quad y=\frac{y^*}{h}, \quad
    y_F = \frac{Y_F}{Y_{F,u}}, \quad \theta=\frac{T-T_u}{T_{ad}-T_u}. \nonumber
\end{align}
and introduce the linear operator
\begin{equation}
    \mathcal{L}_{Le} \equiv \left[\frac{\Pec(1-3y^2)}{2\sqrt{\Dam}} + S\right] \pfr{}{x}- \frac{1}{\Lew}\nabla^2
\end{equation}
where $\nabla^2 = \pfi{^2}{x^2}+\Dam^{-1}\pfi{^2}{y^2} + \pfi{^2}{z^2}$ and the parameters $\Pec$, $\Lew$ and $\Dam$ are as given in
\eqref{NonDimPar}. With these notations, the governing equations take the form
\begin{align}
    \pfr{y_F}{t} +  \mathcal{L}_{Le} y_F =-  \omega,\qquad
    \pfr{\theta}{t} +  \mathcal{L}_1 \theta = \omega \label{yFth}
\end{align}
where  $\mathcal{L}_1= \mathcal{L}_{Le=1}$ and
\begin{equation}
    \omega(y_F,\theta) = \frac{\beta^2y_F}{2\Lew}\exp\left[\frac{-\beta(1-\theta)}{1-\alpha(1-\theta)}\right].
\end{equation}
The corresponding boundary conditions are prescribed below.

\section{Steadily propagating symmetric flames} 
\label{sec:base}

The steady two-dimensional solutions whose stability will be investigated correspond to solutions of \eqref{yFth} which are
independent of $z$ and $t$ and  symmetric with respect
to the $y=0$ plane. These are  denoted by   $y_F=Y(x,y)$ and $\theta=\Theta(x,y)$, and satisfy  the equations
 \begin{align}
   \mathcal{L}_{Le} Y  =- \omega, \qquad
   \mathcal{L}_{1} \Theta  = \omega, \label{YsTs}
\end{align}
subject to the boundary conditions
\begin{align}
    &x\to-\infty: \,\, Y-1=\Theta=0, \\
    &x\to+\infty: \,\, Y=\Theta-1=0,  \label{bc1s}\\
    &y=\pm 1: \,\, \pfr{Y}{y}=\pfr{\Theta}{y}=0 . \label{bc2f}
\end{align}
Since the solutions sought are  symmetric, the problem can be solved on the domain corresponding to the upper-half infinite strip defined by the intervals $x\in(-\infty,\infty)$ and $y\in[0,1]$ using the boundary conditions  
\begin{align}
    \pfr{Y}{y}=\pfr{\Theta}{y}=0\quad \text{at}\quad y=0,1. \label{bc2s}
\end{align}

\begin{figure*}[ht!]
\centering
\includegraphics[width=1.03\textwidth]{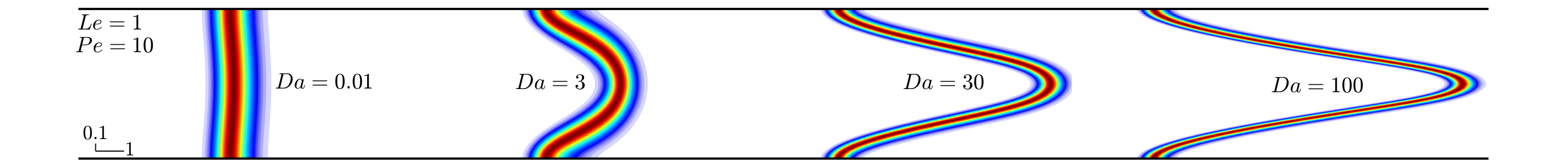}
\includegraphics[width=1.03\textwidth]{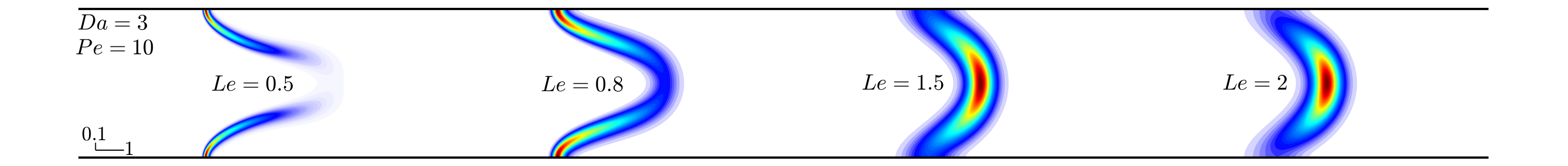}
\includegraphics[width=1.03\textwidth]{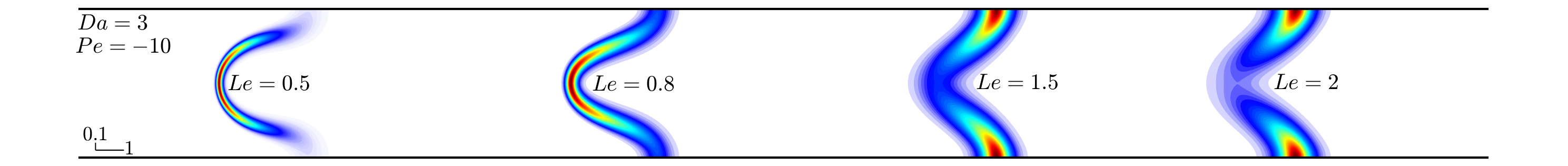}
\caption{Reaction-rate $\omega$-fields for selected values of $\Pec$, $\Lew$ and $\Dam$ with $\beta=10$ and $\alpha=0.85$. The  $x$ and $y$ length scales are indicated in  each subfigure. The unburnt gas is in all cases to the left of the flame.} 
\label{fig:basereact}
\end{figure*}

Selecting $\beta=10$ and $\alpha=0.85$ as representative values for $\beta$ and $\alpha$,   the effective burning speed $S$ and the functions $Y$ and $\Theta$ are determined numerically for given values of $\Lew$, $\Pec$ and $\Dam$. These computations are performed using COMSOL Multiphysics
as described in \cite{daou2023flame, rajamanickam2023thick}. They are based on the finite-element method and use a non-uniform grid with typically 300,000 triangular elements along with local refinement around the reaction zones. A summary of the results is provided in Fig.~\ref{fig:S} and Fig.~\ref{fig:basereact}.

Fig.~\ref{fig:S}(a) shows the  dependence of $S$ on $\Dam$ for selected values of $\Pec$ and $\Lew$, whereas Fig.~\ref{fig:S}(b) displays its dependence on $\Pec$ for $\Lew=1$ and selected values of $\Dam$. As can be seen, the scaled burning speed $S$ exhibits a non-monotonic variation with respect to $\Dam$ for given values of $\Pec$ and $\Lew$. This variation is associated with the fact that for small values of $\Dam$, the diffusion transport and therefore $S$ is influenced by Taylor dispersion, whereas at large values of $\Dam$, $S$ is strongly influenced by curvature effects, as explained in~\cite{rajamanickam2023thick}. 

The trends observed in Fig.~\ref{fig:S}(b) are discussed in~\cite{daou2001flame}. Particularly, the burning speed grows quadratically for small values of $\Pec$ and grows linearly for $|\Pec| \gg 1$ although with different slopes for positive and negative values of $\Pec$, as predicted in~\cite{daou2001flame}.

A sample of the computed results are
presented in Fig.~\ref{fig:basereact} where reaction-rate fields are shown for  selected values of $\Lew$, $\Pec$ and $\Dam$. The top subfigure, corresponding to $\Lew=1$,  illustrates the transition as $\Dam$ is increased  from Taylor dispersion-controlled thick flames to curved thin flames. The middle and bottom  subfigures, pertaining to  $\Lew\neq 1$ and $\Dam=3$, confirm the well-known fact that preferential diffusion and curvature effects determine local burning rates in thin or moderately thick flames. For example,  the local reaction rate is seen to increase when $\Lew<1$   in regions where the flame is convex towards the unburnt gas (the left side of the flame in the figure). At low values of $\Dam$,  such curvature-related effects are in fact negligible  although the flame is still affected by $\Lew$  through Taylor dispersion \cite{rajamanickam2023thick}.

\section{Linear stability problem} 

The linear stability analysis for the steady flames covered in the previous section will now be pursued. To do that, we shall assume
\begin{equation}
\begin{bmatrix}
  y_F \\
  \theta
\end{bmatrix}=
    \begin{bmatrix}
Y(x,y) \\
\Theta(x,y)
\end{bmatrix}+   \begin{bmatrix}
\hat Y(x,y) \\
\hat \Theta(x,y) 
\end{bmatrix}e^{ikz+\sigma t}  \label{normal}
\end{equation}
with $|\hat Y|\ll |Y|$ and $|\hat\Theta|\ll|\Theta|$. Here, $k$ denotes the spanwise wavenumber which is a real number and $\sigma$ is the growth rate of the perturbation, which is in general a complex-valued eigenvalue. The equations governing $\hat Y(x,y)$ and $\hat\Theta(x,y)$, obtained by substituting \eqref{normal} into~\eqref{yFth} and  linearizing, read
\begin{align}
   \left(\mathcal{L}_{Le} + f + k^2/\Lew+\sigma\right) \hat Y &= -g\hat\Theta,  \label{Yp}\\
   \left(\mathcal{L}_1 - g + k^2+\sigma\right) \hat\Theta   &=  f\hat Y, \label{Tp}
\end{align}
where the functions $f(\Theta)$ and $g(\Theta,Y)$ are given by 
\begin{equation}
    f = \frac{g}{\beta Y}[1+\alpha(\Theta-1)]^2 = \frac{\beta^2}{2\Lew}  \exp\left[\frac{\beta(\Theta-1)}{1+\alpha(\Theta-1)}\right]. \nonumber
\end{equation}
The appropriate boundary conditions are given by
\begin{align}
     &x\to\pm\infty: \,\, \hat Y=\hat\Theta=0,  \\
    &y=\pm 1: \,\, \pfr{\hat Y}{y}=\pfr{\hat\Theta}{y}=0. \label{BCy}
\end{align}

The solutions to the eigenvalue problem \eqref{Yp}-\eqref{BCy} fall into two disjoint groups: even solutions (with respect to $y$) for which $\hat Y(x,y)=\hat Y(x,-y)$ and $\hat\Theta(x,y)=\hat\Theta(x,-y)$ and odd solutions for which $\hat Y(x,y)=-\hat Y(x,-y)$ and $\hat\Theta(x,y)=-\hat\Theta(x,-y)$. The problem can thus be solved in the upper-half infinite strip, as done for determining the steady solutions, by imposing the conditions
\begin{align}    
    &y=0: \,\, \pfr{\hat Y}{y}=\pfr{\hat\Theta}{y}=0  \,\, \text{for even solutions},  \label{bcpeven}\\
    &y=0: \,\, \hat Y=\hat\Theta=0 \,\, \text{for odd solutions}. \label{bcpodd}
\end{align}

\section{Indices for eigenfunctions and terminology\label{sec:indices}} 

A  spatial eigenfunction and its eigenvalue  $\sigma$ can be indexed using three numbers $m$, $n$ and $k$ as  
\begin{equation}
    \hat Y_{mn}(x,y) e^{ikz}, \qquad \sigma_{mnk} \equiv \sigma_{mn}(k).  \label{eigenf}
\end{equation}
The spanwise wavenumber $k$ is continuous (since unbounded domain in the $z$-direction is assumed) and lies in the interval $k\in[0,\infty)$. On the other hand, the indices $m$ and $n$, which characterize eigenmodes in the $x$ and $y$ directions, respectively, are discrete and can be ordered such that $m=1,2,3,\dots$ and $n=0,1,2,3\dots$. We note that $n$ represents the number of zero-crossings of the eigenfunctions in the $y$-direction.
It is also important to emphasise that the $n$-modes are arranged such that even solutions correspond to even integer values of $n$ and odd solutions to odd integer values of $n$. For example, when $\Pec=0$, it can be checked that
\begin{align}
    &\hat Y_{mn}(x,y) = \widetilde Y_m(x) \begin{cases}\cos (n\pi y/2), \quad \text{even}\,n \\
    \sin(n\pi y/2), \quad \text{odd}\,n
    \end{cases} \nonumber \\
    &\sigma_{mn}(k) = \widetilde \sigma_{m}(\sqrt{k^2 + n^2\pi^2/4\Dam}) \nonumber
\end{align}
where $\widetilde \sigma_{m}(k)$ and $\widetilde Y_m(x)$ are  eigenvalues and corresponding eigenfunctions  of the classical linear stability problem of the planar flame. The discreteness of
of the index $n$ is due to confinement in the $y$-direction, while the discreteness of $m$, as in the planar flame stability problem,  results from requiring the perturbations to decay  as $x \to \pm \infty$.   Furthermore,  since the even eigenfunctions  do not affect the symmetry of the base solution  about the $y=0$ plane, we shall refer to them as \textit{symmetric} eigenmodes. Similarly, we shall refer to the odd eigenfunctions  as \textit{asymmetric} eigenmodes.

\begin{figure}[ht!]
\centering
\includegraphics[width=0.7\textwidth]{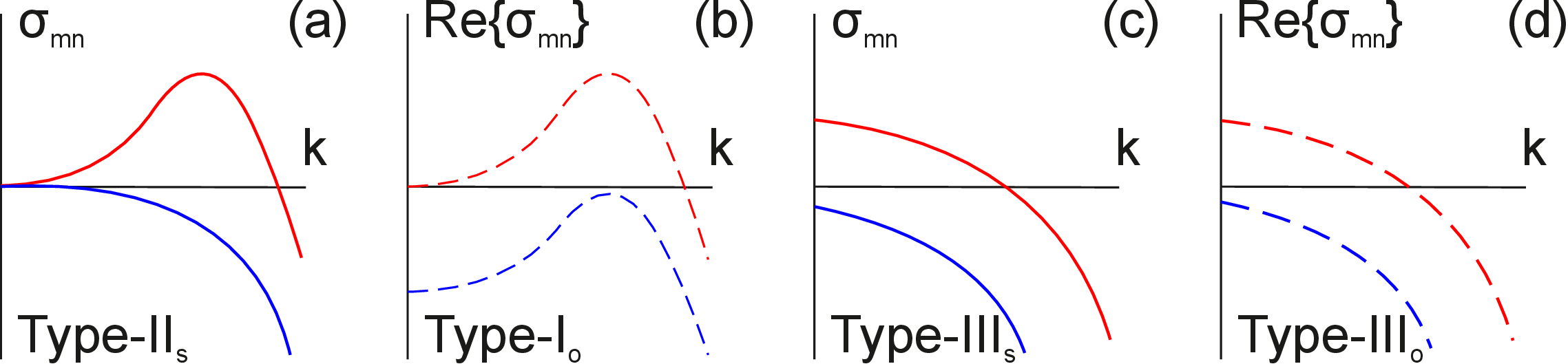}
\caption{\blue{Stable to unstable bifurcation of particular $m$-th dispersion curve $\sigma_{mn}(k)$ as physical parameters are varied, for fixed values of $n$.} Four types of bifurcations are represented, as described in the text. Solid lines pertain to eigenvalues having zero imaginary part, while dashed lines to eigenvalues having a non-zero imaginary part.} \label{fig:DispCurveSch}
\end{figure}

With the notation introduced above, our main task is to obtain the eigenvalues $\sigma_{mn}(k)$ and corresponding eigenfunctions.  These are determined  by solving numerically the two-dimensional eigenvalue problem~\eqref{Yp}-\eqref{bcpodd}   using the eigenvalue solver in COMSOL Multiphysics. In particular, for a given value  of $n$, we \blue{can plot different dispersion curves $\sigma_{mn}(k)$ each corresponding to different values of $m$. These curves undergo changes when relevant parameters such as $\Lew$ are varied. Instability occurs when a portion of any of such dispersion curves corresponds to positive growth rates such that $\mathrm{Re}\{\sigma_{mn}\}>0$.} Four categories of bifurcations from stable to unstable states are obtained in our problem when plotting  $\sigma_{mn}(k)$ as illustrated in  Fig.~\ref{fig:DispCurveSch}. 
Such bifurcations, which are encountered  as a parameter is varied, are referred to as type-II$_s$, type-I$_o$, type-III$_s$ and type-III$_o$ following the terminology used in~\cite{cross1993pattern}. The subscript $s$ denotes stationary or non-oscillatory bifurcation whereas $o$ denotes oscillatory bifurcation. \blue{As can be seen from the schematic illustration shown in Fig.~\ref{fig:DispCurveSch}, type-II$_s$ bifurcation corresponds to a dispersion curve which always passes through the origin in the $k$-$\sigma$ plane whose concavity changes there as a control parameter is varied. On the other hand, type-I$_o$ bifurcation corresponds to a maximum of the dispersion curve crossing the horizontal axis at a wavenumber $k\neq 0$. The other two type-III bifurcations are similar to type-I bifurcation, except that the crossing occurs at $k=0$ (planar mode). Such bifurcations were encountered in recent studies~\cite{daou2024diffusive,daou2023premixed} that were dedicated to the limiting case $\Dam\ll 1$ with a focus on the role played by the direction of flame propagation.}

The classical \textit{\blue{diffusive-thermal} cellular instability} in premixed flames is represented by  type-II$_s$ bifurcation shown in Fig.~\ref{fig:DispCurveSch}(a), whereas the classical \textit{\blue{diffusive-thermal} oscillatory instability} is represented by   type-I$_o$ bifurcation shown in Fig.~\ref{fig:DispCurveSch}(b); \blue{see e.g.~\cite[pp.~477-479]{clavin2016combustion}}. These two instabilities  arise for usual premixed flames in $\Lew<1$ and $\Lew>1$ mixtures, respectively. In addition to these two familiar instabilities, we also come across the types shown in Fig.~\ref{fig:DispCurveSch}(c) and (d) where the most unstable mode occurs at $k=0$ and therefore has no dependence on the $z$-coordinate. Our computations indicate that these type-III bifurcations are linked to eigenmodes  with odd values of $n$ which  lead to asymmetric flames.

\section{Stability regime diagrams} 

\begin{figure*}[h!]
\centering
\includegraphics[scale=0.5]{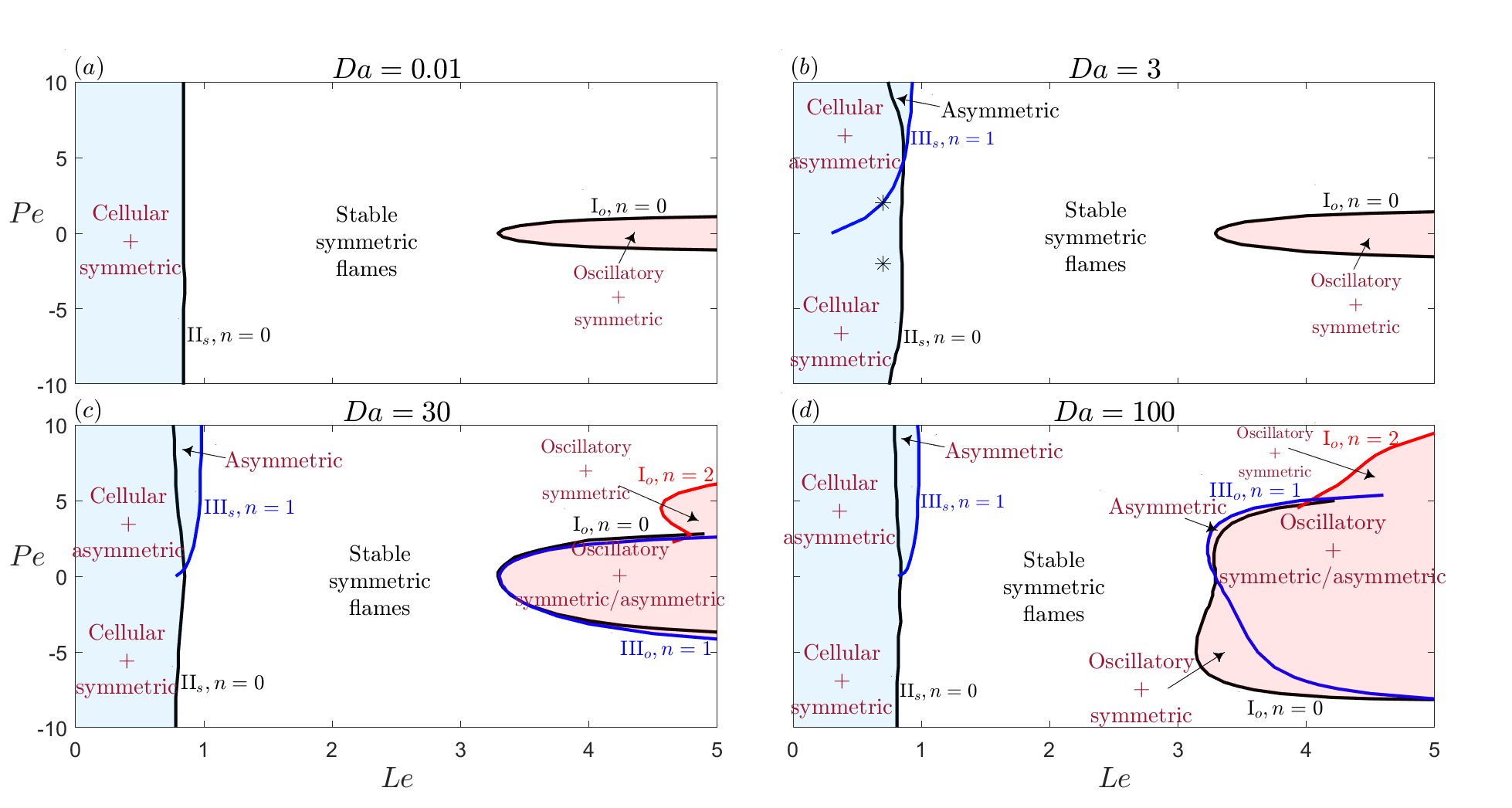}
\caption{Stability regime diagrams in the $\Lew$-$\Pec$ plane for four selected values of $\Dam$. The white regions correspond  to stable, symmetric flames,  whereas the shaded regions to unstable flames. The bifurcation curves are represented by solid lines, and they are labelled by the bifurcations types (I$_o$, II$_s$, III$_s$ and III$_o$) described in Fig.~\ref{fig:DispCurveSch}. The label $n$
refers to the index of the eigenfuction in the $y$ direction defined in section~\ref{sec:indices}.} \label{fig:stability}
\end{figure*}

A convenient way to summarize the findings of the stability analysis is to determine the boundaries of the regions of instability in the $\Lew$-$\Pec$ plane for   fixed values of $\Dam$. This is carried out in Fig.~\ref{fig:stability} for $\Dam=0.01, \, 3, \, 30$ and $100$. The white regions here correspond to stable, symmetric flames, and the shaded regions to unstable flames. The solid lines represent the bifurcation curves and are labelled according to the type of bifurcations as defined in the previous section and in Fig.~\ref{fig:DispCurveSch}.  
Cellular instabilities are encountered for $\Lew<1$ and oscillatory ones for $\Lew>1$. It is also found that  asymmetric flames can appear both in $\Lew<1$ and $\Lew>1$ mixtures.

Fig.~\ref{fig:stability}(a), pertaining to 
$\Dam=0.01$, exhibit the instability regions which occur for thick flames. The cellular instability is found to occur for $\Lew<1$ with its boundary in the $\Lew$-$\Pec$ plane being independent of $\Pec$, although the dispersion curves themselves, similar to those in Fig.~\ref{fig:DispCurveSch}(a)), do depend on  $\Pec$. Drastic changes however happen for $\Lew>1$ mixtures since a slight increase in $\Pec$ is found to completely suppresses the oscillatory instability. These results caused by Taylor dispersion are consistent with the findings reported in~\cite{daou2021effect} for $\Dam\ll 1$. Furthermore, within the range of $\Pec$ and $\Lew$ shown in Fig.~\ref{fig:stability} (a),  asymmetric flames are not encountered and the instability domains appear invariant under the transformation $\Pec\mapsto-\Pec$. 

Fig.~\ref{fig:stability}(b) exhibit the instability regions for $\Dam=3$. Here we can observe that the boundary (left black curve) of the cellular instability region depends on $\Pec$, unlike in the case of Fig.~\ref{fig:stability}(a). More importantly, we now also have a type-III bifurcation associated with the emergence of asymmetric flames for $\Lew<1$ and $\Pec>0$ (blue curve). Therefore, flame propagating against a Poiseuille flow ($\Pec >0$) for subunity Lewis numbers can exhibit a cellular pattern in the spanwise $z$-direction and an asymmetric pattern in the $y$-direction which is normal to the wall. However, if the flame propagation is aided by the  flow ($\Pec <0$), then only cellular  symmetric flames can arise. The direction of flame propagation relative to the flow direction is therefore an important factor in determining the emergence of asymmetric flames.

Fig.~\ref{fig:stability}(c) pertains to the moderately large value $\Dam=30$, and shows that the instability regions for $\Lew<1$ are qualitatively similar to those in Fig.~\ref{fig:stability}(b). Here the asymmetric flames are more readily accessible as they are obtained even for smaller values of $\Pec$. A noteworthy observation is that  asymmetric flames can now also occur in $\Lew>1$ mixtures. Furthermore, in addition to the $n=0$   and $n=1$   eigenmodes, we also observe $n=2$  eigenmodes (red curve) when $\Lew>1$  and $\Pec >0$.

Fig.~\ref{fig:stability}(d), pertaining to $\Dam=100$ show that the instability region possesses qualitatively similar characteristics to that in Fig.~\ref{fig:stability}(c) in the  $\Lew<1$ domain. The instability region in the $\Lew>1$ domain is however greatly enlarged and relatively more complex.  In particular, the instability region moves to the left, towards smaller values of  
$\Lew$, most notably for $\Pec<0$.
This makes this instability more easily accessible. 

At this stage, it is worth comparing our results with those reported by Kurdyumov~\cite{kurdyumov2011lewis}, which do not account for perturbations in the $z$-direction, and involve in fact only $k=0$ modes and conditions corresponding to $\Pec\geq 0$. It is clear from Fig.~\ref{fig:DispCurveSch} that his analysis  correctly predicts the type-III bifurcations identified herein  since they   correspond to situations where the most unstable mode occurs indeed at $k=0$. In other words, the findings of \cite{kurdyumov2011lewis} give  correct predictions about the appearance of asymmetric flames,
which correspond to the blue curves in Fig.~\ref{fig:stability}. These findings do not provide however
the boundaries characterising the onset of the cellular and oscillatory instabilities (black and red curves in Fig.~\ref{fig:stability}), which are determined by dominant unstable modes with $k \neq 0$.

\section{Two-dimensional time-dependent numerical simulations}

\begin{figure*}[ht!]
\centering
\includegraphics[width=1\textwidth]{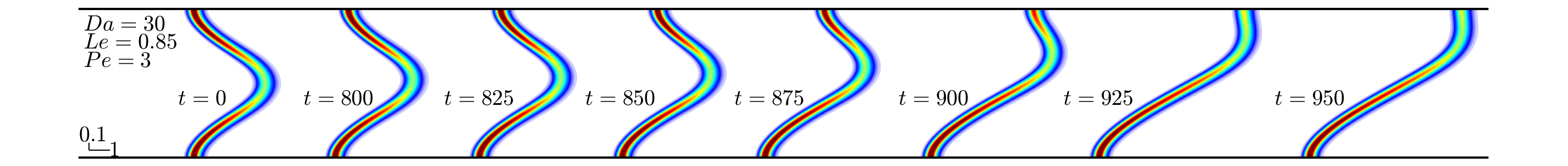}
\includegraphics[width=1\textwidth]{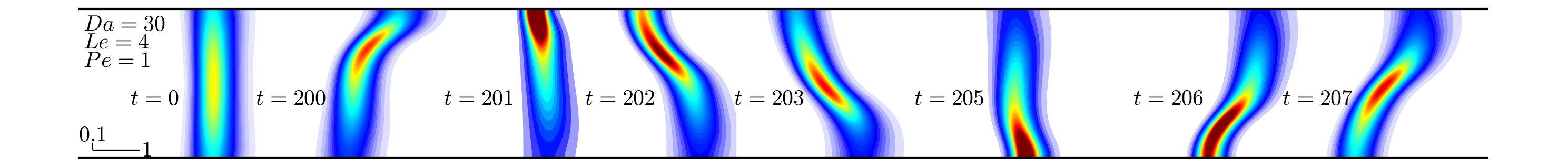}
\includegraphics[width=1\textwidth]{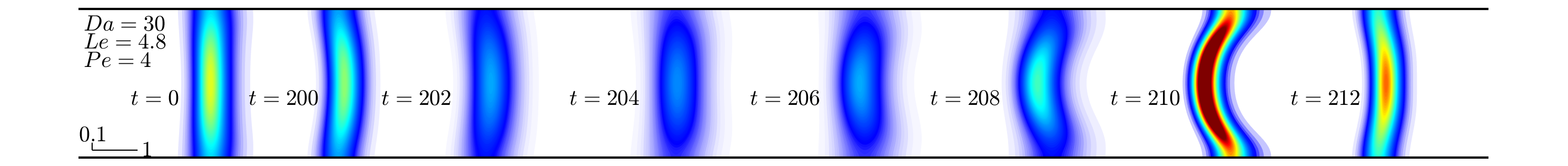}
\caption{Reaction rate $\omega$-fields at selected times, computed for $\beta=10$, $\alpha=0.85$, $\Dam=30$ and selected values of $\Pec$ and $\Lew$. The $x$ and $y$ length scale are indicated at the left bottom of each subfigure.} \label{fig:2Dwplots}
\end{figure*}

To complement  the comprehensive findings of the linear stability analysis summarised in Fig.~\ref{fig:stability},   time-dependent simulations have been performed to determine the evolution of unstable symmetric flames.  Two-dimensional simulations are discussed in this section and three-dimensional ones in the following section. The two-dimensional simulations  are carried out in the $x$-$y$ domain $[-50,50]\times[-1,1]$, with initial conditions  corresponding to steady symmetric flames described in section~3.

Computed reaction rate $\omega$-fields are shown in Fig.~\ref{fig:2Dwplots} for $\Dam=30$ and selected values of $\Pec$, $\Lew$ and time $t$. The top subfigure, pertaining to $\Lew=0.85$ and $\Pec=3$, illustrates the transition from symmetric to asymmetric flames. For this case, the  transition results ultimately in a steadily-propagating stable asymmetric flame, whose effective burning speed is found to be higher  than that of the \blue{ symmetric flame~\cite{kurdyumov2011lewis,rodriguez2023characterization}.} The middle subfigure, pertaining $\Lew=4$ and $\Pec=1$,  also exhibits  the aforementioned transition, although it leads ultimately  to a quasi-periodic, rather than a steady, solution. The bottom subfigure, pertaining to $\Lew=4.8$ and $\Pec=4$, shows that the flame  evolution involves persistent quasi-periodic oscillations, with the solution remaining however symmetric, a behaviour which  is consistent with the predictions of the linear stability analysis in Fig.~\ref{fig:stability}(c) (in the region with the label ``oscillatory + symmetric").

\section{Three-dimensional time-dependent simulations} 

\begin{figure}[ht!]
\centering
\includegraphics[width=0.49\textwidth]{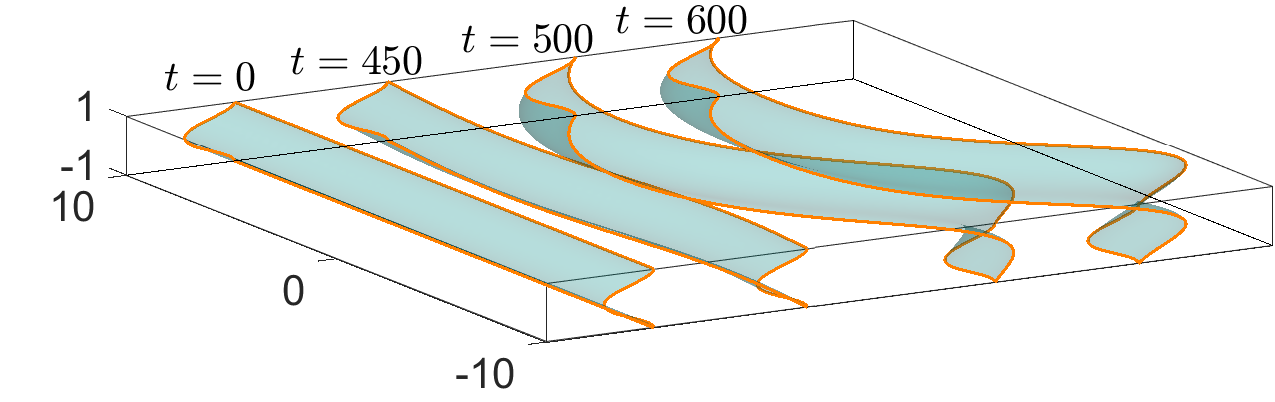}
\includegraphics[width=0.49\textwidth]{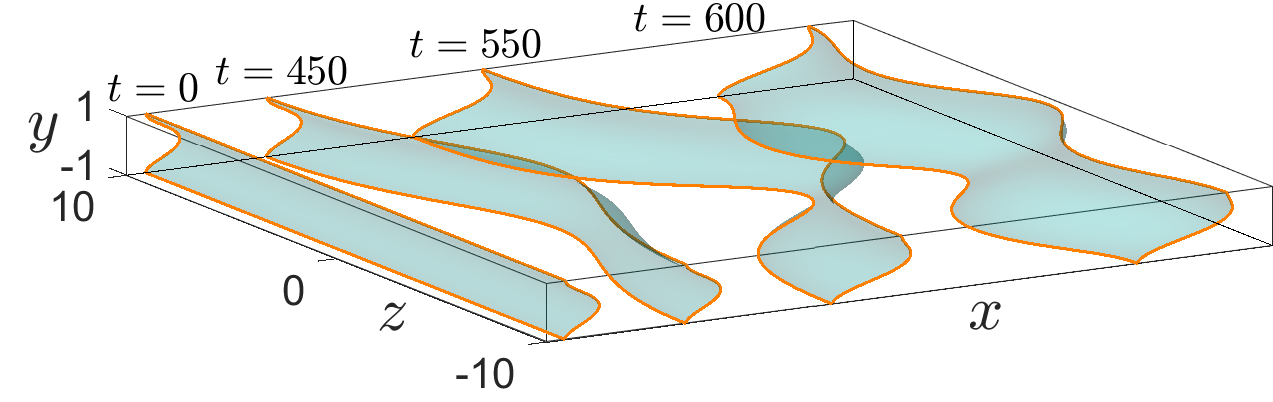}
\caption{Time evolution of the flame,  represented by the iso-surface $\theta=0.9$, for $\Dam=3,\ \Lew=0.7,\ \Pec=-2$ (top subfigure) and $\Pec=2$ (bottom subfigure).} \label{fig:TD3D}
\end{figure}

\begin{figure*}[ht!]
\centering
\includegraphics[width=1\textwidth]{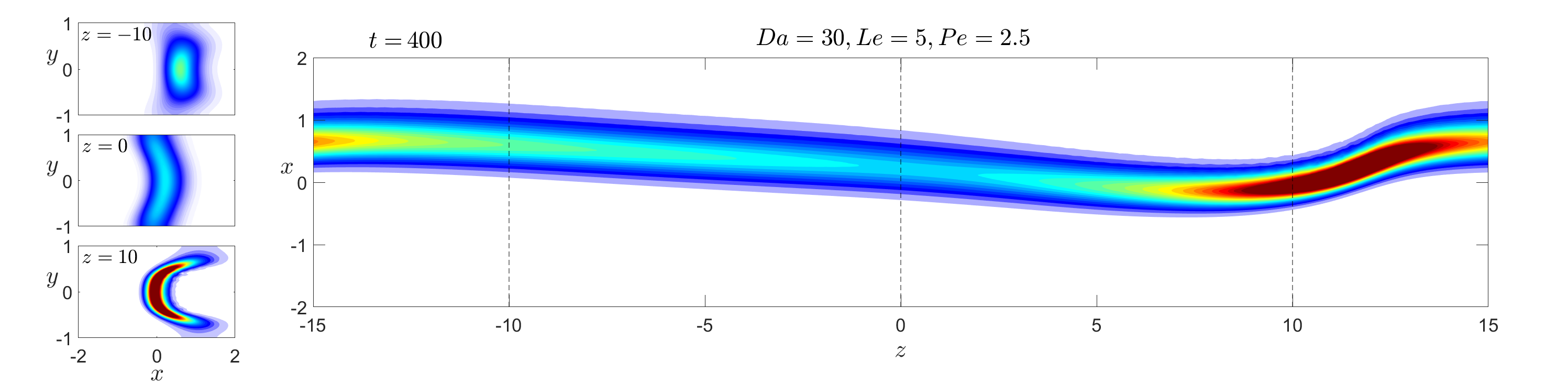}
\caption{Reaction rate $\omega$-fields at a selected value of time, $t=400$, computed for $\beta=10$, $\alpha=0.85$, $\Dam=30$, $\Pec=2.5$ and $\Lew=5$. The three subfigures on the left correspond to sections of the domain by the planes $z=-10$, $z=0$ and $z=10$ whereas the right subfigure corresponds to  a section by the  plane $y=0$.} \label{fig:Da30}
\end{figure*}

The two-dimensional simulations discussed in the previous section are particularly significant    in cases where the most unstable mode occurs at $k=0$; such cases belong to the regions labelled ``asymmetric" in Fig.~\ref{fig:stability}. For situations where the most unstable mode does not occur  at $k=0$, three-dimensional simulations are important. Computations for three sample cases are presented in Fig.~\ref{fig:TD3D} and Fig.~\ref{fig:Da30} which use periodic boundary conditions in the $z$-direction. 
 These computations are carried out using COMSOL Multiphysics, as for the two-dimensional simulations,  and are performed on non-uniform tetrahedral grids.

The iso-surface $\theta=0.9$, taken as a representative of the flame surface, is shown in Fig.~\ref{fig:TD3D} at selected values of time $t$. The top subfigure,  corresponding to $\Lew=0.7$, $\Pec=-2$ and $\Dam=3$, illustrates the formation of a cellular pattern in the spanwise $z$-direction, while remaining symmetric at all times about the   $y=0$ plane. In this case, the flame is found to evolve in time into a stable steady state.  The bottom subfigure, corresponding to $\Lew=0.7$, $\Pec=2$ and $\Dam=3$,  demonstrates the formation of a cellular pattern in the  $z$-direction, but also the transition to asymmetric flames. In this case, however, the flame is not found to approach a steady state. The complete flame dynamics for these two cases can be better appreciated from the animated time evolution included in the supplementary material.



Computational results for $\Lew=5$, $\Pec=2.5$ and $\Dam=30$ are presented in Fig.~\ref{fig:Da30}. Shown are the reaction rate $\omega$-fields in the $y=0$ plane (top view) and in  the  planes  defined by $z=-10$, $z=0$ and $z=10$, at a particular value of time, $t=400$. The flame evolves
into a time-dependent structure involving traveling waves along the $z$-direction  and  time oscillations of a symmetric flame in planes parallel to the $x$-$y$ plane. Please refer to the supplementary material for a   time animation of this case.


\section{Conclusions}

In this study,   the diffusive-thermal instability of flames propagating in a Poiseuille flow is investigated  within a three-dimensional framework. A wide range of variation of the physical parameters is considered which covers a variety of practical applications \blue{ranging from sub-millimetre micro-burners to relatively large channel burners} and experiments involving flame propagation in  wide  or narrow channels \blue{and phenomena such as flame flashbacks}. 

The flame is found, depending on the parameters,  to undergo either an oscillatory instability, or a cellular instability in the spanwise direction (the $z^*$-direction in Fig.~\ref{fig:setup}) combined with a transition instability from symmetric  to asymmetric shapes in planes parallel to the $x^*$-$y^*$ plane. The instabilities encountered are determined by    the system  size (or, the Damk{\"o}hler number $\Dam$), the flow strength (or, the Peclet number $\Pec$) and by preferential diffusion  (or, the Lewis number $\Lew$). The results of a linear stability problem, based on the computation of its eigenvalues,  are conveniently summarized in the form of stability regime diagrams drawn in the $\Lew$-$\Pec$ plane for selected values of $\Dam$, see Fig.~\ref{fig:stability}.   Two- and three-dimensional time-dependent computations are also carried out, confirming the prediction of the linear stability analysis for sample values of the parameters, and describing the full evolution of unstable  flames  into the non-linear regime.

In narrow channels, it is found that the flames are always symmetric about the channel mid-plane and can be strongly influenced by Taylor dispersion;  in particular, Taylor dispersion is found to promote  the cellular instability  and to inhibit the oscillatory instability. In a moderate size or large  channels, on the other hand, the flame has  stronger propensity for $\Lew<1$ to become asymmetric    when it is opposed by the flow ($\Pec >0$), but not when it is aided by the flow ($\Pec <0$). In addition, the instability scenarios for $\Lew>1$ are more complex and involve  both symmetric and asymmetric flames; see e.g. Fig.~\ref{fig:stability}(c) and (d).

\blue{Since the conclusions drawn above are based on the diffusive-thermal model under adiabatic conditions, it is desirable to extend this investigation to account for thermal-expansion and heat-loss effects in the future. For instance, in the presence of heat losses, the boundaries of the instability domains drawn in Fig.~\ref{fig:stability} are anticipated to shift towards $\Lew=1$~\cite{joulin1979linear}, making the instability more readily accessible. In the presence of density or viscosity variations, in addition to the diffusive-thermal instability addressed herein, hydrodynamic instabilities such  as the Darrieus--Landau  and the Saffman--Taylor instabilities are also expected to contribute to the instability of flames~\cite{joulin1994influence,han2021effect,fernandez2023three}.}

\section*{Declaration of competing interest}

The authors declare that they have no known competing financial interests or personal relationships that could have appeared to influence the work reported in this paper.

\section*{Acknowledgments} 

This work was supported by the UK EPSRC through grant EP/V004840/1

\section*{Supplementary material} 

Animated time evolution of numerical simulations are included as video files.

\bibliographystyle{elsarticle-num}

\bibliography{elsarticle-template}

\end{document}